\documentclass[11pt,twoside]{article}

\usepackage{asp2014}

\aspSuppressVolSlug
\resetcounters

\begin{document}

\title{Building an interoperable, distributed storage and authorization system}

\author{S.~Bertocco,$^1$ B.~Major,$^2$ P.Dowler,$^2$ S.Gaudet,$^2$ M.Molinaro,$^1$ and G.Taffoni$^1$}
\affil{$^1$, INAF-OATs, Istituto Nazionale di Astrofisica - Osservatorio Astronomico di Trieste, Trieste, Italy}
\affil{$^2$CADC, National Research Council Canada, Victoria, BC, Canada}

\paperauthor{Sara~Bertocco}{bertocco@oats.inaf.it}{orcid.org/0000-0003-2386-623X}{Istituto Nazionale di Astrofisica}{Osservatorio Astronomico di Trieste}{Trieste}{}{34143}{Italy}
\paperauthor{Brian~Major}{majorb@nrc-cnrc.gc.ca}{0000-0001-9263-3048}{National Research Council Canada}{CADC}{Victoria}{British Columbia}{}{Canada}
\paperauthor{Patrick~Dowler}{patrick.dowler@nrc-cnrc.gc.ca}{orcid.org/0000-0001-7011-4589}{National Research Council Canada}{CADC}{Victoria}{British Columbia}{}{Canada}
\paperauthor{Séverin~Gaudet}{Severin.Gaudet@nrc-cnrc.gc.ca}{orcid.org/0000-0001-9281-2361}{National Research Council Canada}{CADC}{Victoria}{British Columbia}{}{Canada}
\paperauthor{Marco~Molinaro}{molinaro@oats.inaf.it}{orcid.org/0000-0001-5028-6041}{Istituto Nazionale di Astrofisica}{Osservatorio Astronomico di Trieste}{Trieste}{}{34143}{Italy}
\paperauthor{Giuliano~Taffoni}{taffoni@oats.inaf.it}{orcid.org/0000-0002-4211-6816}{Istituto Nazionale di Astrofisica}{Osservatorio Astronomico di Trieste}{Trieste}{}{34143}{Italy}

\begin{abstract}
A joint project between the Canadian Astronomy Data Centre (CADC) of the Canadian National
Research Council, and the Italian Istituto Nazionale di Astrofisica-Osservatorio
Astronomico di Trieste (INAF-OATs), partially funded by the EGI-Engage
H2020 European Project, is devoted to deploy an integrated infrastructure
to access and exploit astronomical data. This infrastructure will be entirely based on the 
International Virtual Observatory Alliance (IVOA) standards, see References~\citet{ivoa}.
Currently CADC-CANFAR provides scientists with an access, storage and computation
facility based on software libraries implementing a set of standards and recommendations developed
by the International Virtual Observatory Alliance (IVOA).
The deployment of a twin infrastructure, basically built on the same open source
software libraries, has been started at INAF-OATs. Currently, this new infrastructure 
provides users with an Access Control Service and a Storage Service. The final
goal of the ongoing project is to build an integrated infrastructure
providing complete interoperability between the two described geographically 
distributed infrastructures, both in users access control and data sharing.
This paper describes the target infrastructure, the main user requirements
covered, the technical choices and implemented solutions.

\end{abstract}

\section{IVOA Standards}

The Virtual Observatory (VO) is the vision that astronomical datasets and other 
resources should work as a seamless whole, exploitable in a single 
transparent system.  

The International Virtual Observatory Alliance (IVOA) is an organisation that 
debates and agrees the technical standards needed to make the VO possible. 
It is a framework for discussing and sharing VO ideas and technology, and a body 
for promoting and publicising the VO.

The infrastructure we are describing is based on IVOA standards/recommendations 
because their implementation ensures the interoperability of all the  
geographically distributed services.
 
In particular the implemented standards/recommendations are:

\textbf{VOSpace 2.1}: \textbf{VOSpace} is the IVOA interface to distributed storage.

\textbf{SSO 1.01}: \textbf{IVOA Single-Sign-On Profile} describes approved 
client-server authentication mechanisms  

\textbf{CDP 1.0}: the \textbf{Credential Delegation Protocol} allows a client 
program to delegate a user's credentials to a service such that that service 
may make requests of other services in the name of that user.

\textbf{UWS 1.1}: the \textbf{Universal Worker Service Pattern} defines how 
to manage asynchronous execution of jobs on a service.  

\textbf{VOSI 1.1}: \textbf{IVOA Support Interfaces} describes the minimum 
interface that a  web service requires to participate in the IVOA, i.e. a 
set of common basic functions that all these services should provide in 
the form of a standard support interface in order to support the effective 
management of the VO.

An implementation of this standards/recommendations can be found on a
GitHub repository, see References~\citet{opencadcrepo}.

\section{Twin infrastructures to access and exploit astronomical data}

The Canadian Advanced Network for Astronomical Research (CANFAR)
provides to its users access to very large resources for both
storage and processing, using a cloud based framework, see References~\citet{infrastructure}. 
This infrastructure
is widely used and appreciated and the software facilities available on top of it
are based on a set of APIs freely available under the terms of the GNU Affero
General Public License as published by the Free Software Foundation, and
published as github repository, see References~\citet{opencadcrepo}. 
Founding on these reasons,
at INAF-OATs it is a work in progress to deploy a twin framework,
with the same technical requirements and use cases and based on the same software
libraries. At present this new infrastructure provides to its users
an authentication and authorization framework and a data storage and
managements service homogeneous with the analogous CANFAR services.

\section{Access Control}

The first step to integrate and make interoperable the two twin
infrastructures for data storage and processing, described above,
is the achievement of the authentication and authorization services
interoperability.
In each infrastructure framework, the Access Control service provides
authentication and identity management
support to web services or to clients directly.
The authorization service supports multiple identities (i.e.
login-password, x509 certificates, cookies based authentication): users
may have multiple identities
and can connect to the ecosystem of services with any of those identities.
The Access Control is based on the group membership concept. Users can be
members, administrators, or owners of groups.  Ownership and
administrative membership allows for different levels of group management.
Users are considered to be authorized to a resource (a service or proprietary
data, for example) if they are a member of the group(s) protecting that
resource.  Resource protection is achieved by the owner of the resource
assigning (granting) a group to that resource.
The Access Control implementation is based on the native
Java Authentication and Authorization Service (JAAS) APIs and it exposes
a RESTful interface.
The Access Control interoperability between the two infrastructures is based
on the IVOA credential delegation protocol specification exploitation: both
the infrastructures are provided with a credential delegation service,
the user of one infrastructure gains access to it querying its access
control service, he delegates his credentials to the credential delegation
service of the second infrastructure and access it querying its access
control service using his delegated identity.
\articlefigure{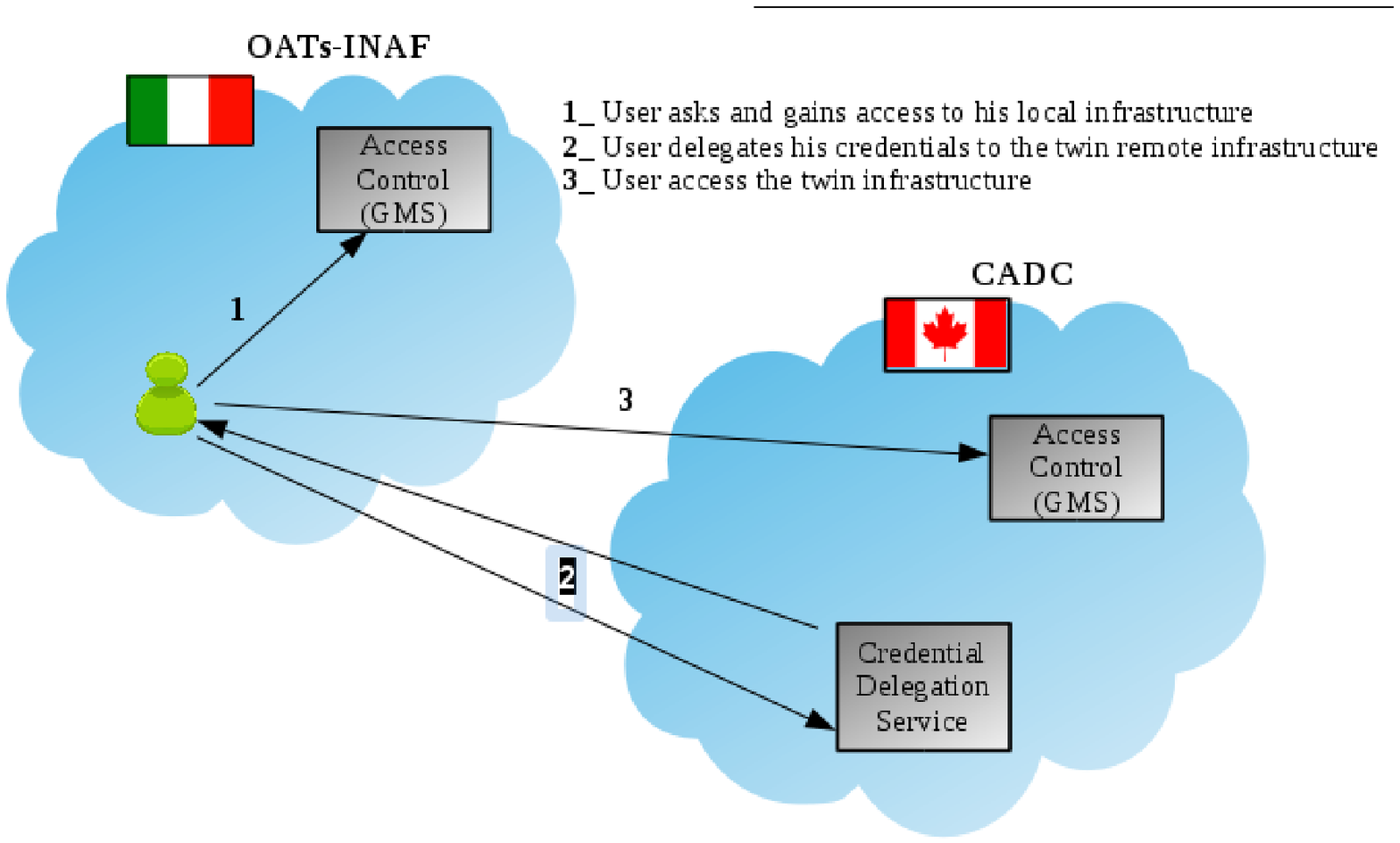}{ac_interop}{Access Control interoperability.}

\section{VOSpace}
The Storage Service is built of two main components: a VOSpace front-end,
implementing the IVOA 2.1 VOSpace standard and a back-end, managing the physical
data storage and retrieval.
The front-end in both the infrastructures is based on the open source
implementation, provided by CADC, of the IVOA 2.1. It exposes a RESTful
interface and it manages the user requests, the user authentication
and authorization verification, the metadata relative to the stored resources.
The back-end is a pluggable component and it is differently developed in the
two infrastructures: it is proprietary file system based at CADC
and posix based at OATs-INAF. In both cases it exposes a RESTful interface.
The main architectural concepts applied are: RESTful architecture to
achieve an easy to manage services distribution
and plugins architecture to allow for a smooth technology substitution.
The separation of the two components is preserved to allow the use of
different back-ends with a minimal effort.
\articlefigure{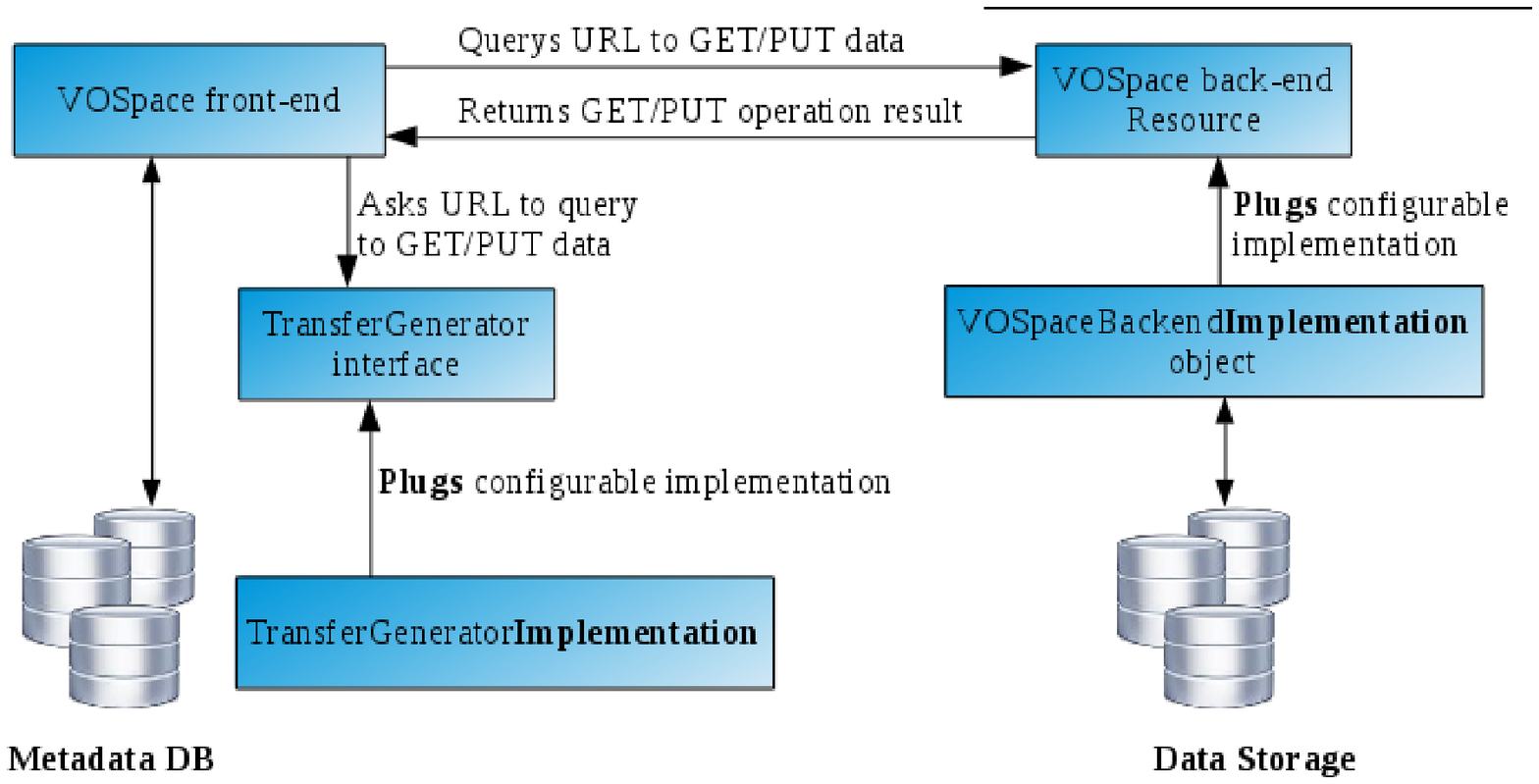}{vospace_architectural_concepts}{VOSpace architectural base.}
Figure \ref{vospace_architectural_concepts} shows the VOSpace service
main architectural features: the VOSpace front-end receives user requests,
manages the relative metadata, asks a TransferGenerator object to obtain
the URL to use to query the back-end. The TransferGenerator object is pluggable,
so different services can be queried. The VOSpace front-end now is able to query
the back-end to access the data. The back-end implementation is also pluggable,
so different storage implementations can be used.

\section{Future plans}
This framework interoperability will be extended integrating the
italian infrastructure into the EGI cloud. This way the Canadian side will be
reached by European users and vice versa.

The VOSpace storage back-end, now realized as a posix-based file system, just 
to easy and fast obtain a proof of concept, will be added with other plugins
to more efficient storage systems. OpenStack Swift, CEPH and OneData are under 
evaluation.

\bibliography{poster_P4.2}  

\end{document}